\documentclass[pdflatex,sn-vancouver-num, referee]{sn-jnl}

\usepackage[utf8]{inputenc}
\usepackage[T1]{fontenc}
\usepackage{graphicx}%
\usepackage{multirow}%
\usepackage{amsmath,amssymb,amsfonts}%
\usepackage{amsthm}%
\usepackage{mathrsfs}%
\usepackage[title]{appendix}%
\usepackage{xcolor}%
\usepackage{textcomp}%
\usepackage{manyfoot}%
\usepackage{booktabs}%
\usepackage{algorithm}%
\usepackage{algorithmicx}%
\usepackage{algpseudocode}%
\usepackage{listings}%

\begin{document}

\title[Improving motor imagery decoding methods for an EEG-based mobile brain-computer interface in the context of the 2024 Cybathlon]{Improving motor imagery decoding methods for an EEG-based mobile brain-computer interface in the context of the 2024 Cybathlon}

\author[1,3]{\fnm{Isabel W.} \sur{Tscherniak}}
\equalcont{These authors contributed equally to this work.}
\email{team@neurotum.com}

\author[1,3]{\fnm{Niels C.} \sur{Thiemann}}
\equalcont{These authors contributed equally to this work.}
\email{team@neurotum.com}

\author[1,3]{\fnm{Ana} \sur{McWhinnie-Fernández}}
\equalcont{These authors contributed equally to this work.}
\email{team@neurotum.com}

\author[1,3]{\fnm{Iustin} \sur{Curcean}}
\author[2,3]{\fnm{Leon L.J.} \sur{Jokinen}}
\author[2,3]{\fnm{Sadat} \sur{Hodzic}}
\author[1,3]{\fnm{Thomas E.} \sur{Huber}}
\author[1,3]{\fnm{Daniel} \sur{Pavlov}}
\author[1,3]{\fnm{Manuel} \sur{Methasani}}
\author[3]{\fnm{Pietro} \sur{Marcolongo}}
\author[1,3]{\fnm{Glenn V.} \sur{Krafczyk}}
\author[1,3]{\fnm{Oscar Osvaldo} \sur{Soto Rivera}}
\author[3]{\fnm{Thien} \sur{Le}}
\author[3]{\fnm{Flaminia} \sur{Pallotti}}
\author[1,3]{\fnm{Enrico A.} \sur{Fazzi}}

\affil[1]{\orgname{Technical University of Munich}, \orgaddress{\street{Arcisstrasse 21}, \postcode{80333} \city{Munich}, \country{Germany}}}

\affil[2]{\orgname{Ludwig-Maximillian University of Munich}, \orgaddress{\street{Geschwister-Scholl-Platz 1}, \postcode{80539} \city{Munich}, \country{Germany}}}

\affil[3]{\orgname{neuroTUM e.V.}, \orgaddress{\street{Karlstrasse 45}, \postcode{80333} \city{Munich}, \country{Germany}}}

\abstract{\textbf{Background:} Brain-computer interfaces (BCIs) have shown significant promise over the past decades, but often fail to meet the requirements for portability and usability in real-world scenarios. Motivated by the Cybathlon 2024 competition, we developed a modular, online EEG-based BCI system to address these challenges, thereby increasing accessibility for individuals with severe mobility impairments, such as tetraplegia.
\par
\textbf{Methods:} Our system uses three mental and motor imagery (MI) classes to control up to five control signals. The pipeline consists of four modules: data acquisition, preprocessing, classification, and the transfer function to map classification output to control dimensions. These modules run in parallel to optimise the online delay. The preprocessing includes a linear phase filter bandpass, artefact removal, and epoching with a sliding time window. The feature extraction was done using Morlet wavelets and the common spatial pattern. As our deep learning classifier, we use three diagonalized structured state-space sequence layers.
We developed a training game for our pilot where the mental tasks control the game during quick-time events. We implemented a mobile web application for live user feedback. The components were designed with a human-centred approach in collaboration with the tetraplegic user.
\par
\textbf{Results:} We achieve up to 84\% classification accuracy in offline analysis using an S4D-layer-based model. In the live Cybathlon competition setting, our pilot successfully completed one task; we attribute the reduced performance in this context primarily to factors such as stress and the challenging competition environment.
Following the Cybathlon, we further validated our pipeline with the original pilot and an additional participant, achieving a success rate of 73\% in real-time gameplay. 
We also compare our model to the EEGEncoder, which is slower in training but has a higher performance.
The S4D model outperforms the reference machine learning models.
\par
\textbf{Conclusions:} We provide insights into developing a framework for portable BCIs, taking a step towards bridging the gap between the laboratory and daily life. Specifically, our framework integrates modular design, real-time data processing, user-centred feedback, and low-cost hardware to deliver an accessible and adaptable BCI solution, addressing critical gaps in current BCI applications.

}






\maketitle

\section{Introduction}\label{intro}
Brain-computer interfaces (BCIs) enable direct communication between the brain and external devices by decoding neural activity into control commands \cite{Slutzky2025_BCIworkingdefinition}. These systems offer transformative potential for individuals with severe motor impairments, providing new avenues for restoring autonomy and enabling control of assistive technologies such as robotic prostheses, wheelchairs, or virtual environments. Among the various modalities available, non-invasive electroencephalography (EEG) remains the most widely used due to its safety, portability, and relative affordability.
\par
Over the past two decades, significant advances in signal processing and machine learning have improved the performance of EEG-based BCIs \cite{KHADEMI2023109736}. However, major challenges persist that hinder their translation from research laboratories to real-world use. BCIs must maintain reliability under variable conditions, adapt to user-specific neural patterns, and operate robustly over time-dependent data shifts \cite{RAZA2019154}. Moreover, real-time performance, user comfort, and the integration of feedback mechanisms remain critical factors affecting their long-term usability. Addressing these issues requires systems that are not only technically accurate but also practical, modular, and designed with the end user’s needs in mind \cite{Kuebler2014}, \cite{KleihKuebler2018_UCD_BCI}.
\par
The Cybathlon Competition\footnote{See https://www.cybathlon.com (last accessed November 12, 2025)}, organized by ETH Zurich every four years, provides an international platform to evaluate and accelerate the development of assistive technologies in realistic settings. During the competition, individuals with physical disabilities, referred to as "pilots", compete in tasks that simulate everyday challenges using advanced assistive devices, including BCIs, robotic prostheses, and powered exoskeletons \cite{Jaeger2023}. In the BCI race, pilots with tetraplegia control avatars in a virtual game environment using invasive and noninvasive BCIs. 
\par
The BCI is built on the principle that the pilots can select commands by generating specific brain signals. A common type used is motor imagery (MI), which involves imagining motor movements, such as those of the upper and lower limbs, without actually performing them \cite{Khan2020_MI_BCI_Stroke}. Neural responses from MI have been shown to be highly comparable to those elicited by actual movements \cite{Jeannerod1995_MI}, \cite{Kilner2004_MI}.
MI relies on internal activity, which is non-stimulus-dependent, and represents the activation of the sensorimotor cortex when the user imagines physical movement \cite{Lotze2006_MI}. For this reason, we selected MI as the primary control strategy throughout the development of the BCI pipeline. The system was designed to produce three continuous control signals and two binary control signals.
\par
Motivated by the Cybathlon 2024, we developed a modular, mobile EEG-based BCI system tailored to our pilot, but which can be easily adapted to other users. Our system was designed to address key challenges of portability, low latency, and user-centered adaptability while maintaining competitive decoding performance. The architecture integrates distinct modules for data acquisition, preprocessing, classification, and feedback, operating in both offline and online modes to optimize real-time performance.
Despite rapid advances in BCI technology, achieving robust and reliable control in real-world environments, such as the dynamic and high-pressure setting of the Cybathlon, remains a major challenge. Leveraging recent advances in deep learning, we implement a classifier based on the linear recurrent neural network diagonalized Structured State-Space Sequence (S4D) layer to enhance classification accuracy and signal stability, which is novel to motor imagery classification. 
\par
A crucial aspect of our approach is the incorporation of human-centered design principles throughout the development process. Working closely with our tetraplegic pilot, we iteratively refined the selection of mental tasks and feedback design. Additionally, we developed a game-based training environment to enhance engagement, comfort, and consistency of control. By emphasizing usability and adaptability, our system aims to bridge the gap between high-performing laboratory BCIs and accessible assistive solutions for everyday use.
\par
In this paper, we present the design, implementation, and evaluation of a motor imagery-based BCI developed for the 2024 Cybathlon competition. The system integrates a human-centered design framework with a robust signal processing and classification pipeline, enabling intuitive, reliable, and user-adaptive control. We describe the offline training procedures, real-time architecture, and interactive feedback modules that support adaptability between the user and decoder. Experimental results demonstrate the system’s accuracy, responsiveness, and low-latency performance under real-world conditions. Finally, we reflect on the insights gained from deploying the system in a competitive setting and discuss the implications for the future development of portable, user-centered, and explainable BCI systems. The main contributions of this work are as follows:
\begin{enumerate}
    \item Human-centered system design: Integration of user feedback and iterative adaptation throughout the development of the BCI interface, feedback, and task paradigms.
    \item Closed-loop training environment: Development of an interactive game-based training paradigm that maintains engagement while enabling precise monitoring and labeling of motor imagery performance.
    \item Robust online processing pipeline: Implementation of a real-time architecture for EEG decoding, featuring artifact correction, adaptive re-referencing, and stable classification performance.
    \item Real-world validation: Evaluation of system accuracy, responsiveness, and user adaptability during the 2024 Cybathlon competition, providing insights into BCI performance under real-world conditions.
\end{enumerate}

\section{Methods}\label{methods}
\subsection{Human-centered design}
A human-centered design approach was adopted throughout this work. The BCI system, user interfaces, and game environments were developed in collaboration with the pilot to enhance usability and engagement. Both the selection of the mental tasks and the design of the feedback screen were carried out iteratively, incorporating continuous input from the team's pilot.
By directly involving users in the selection and iterative refinement of mental tasks, BCIs can better support adaptability, reduce cognitive workload, and enhance overall performance \cite{Tzimourta2025_HCD_DigitalHealth}. A tailored BCI system also enhances user comfort, enabling longer sessions and facilitating the collection of larger datasets. Furthermore, this approach helps address variability in user performance by adapting the system to individual preferences and capabilities \cite{Kuebler2014}, \cite{Herbert2024_LanguageCultureBCI}.

\subsection{Task design}
\begin{figure}
    \centering
    \includegraphics[width=1\linewidth]{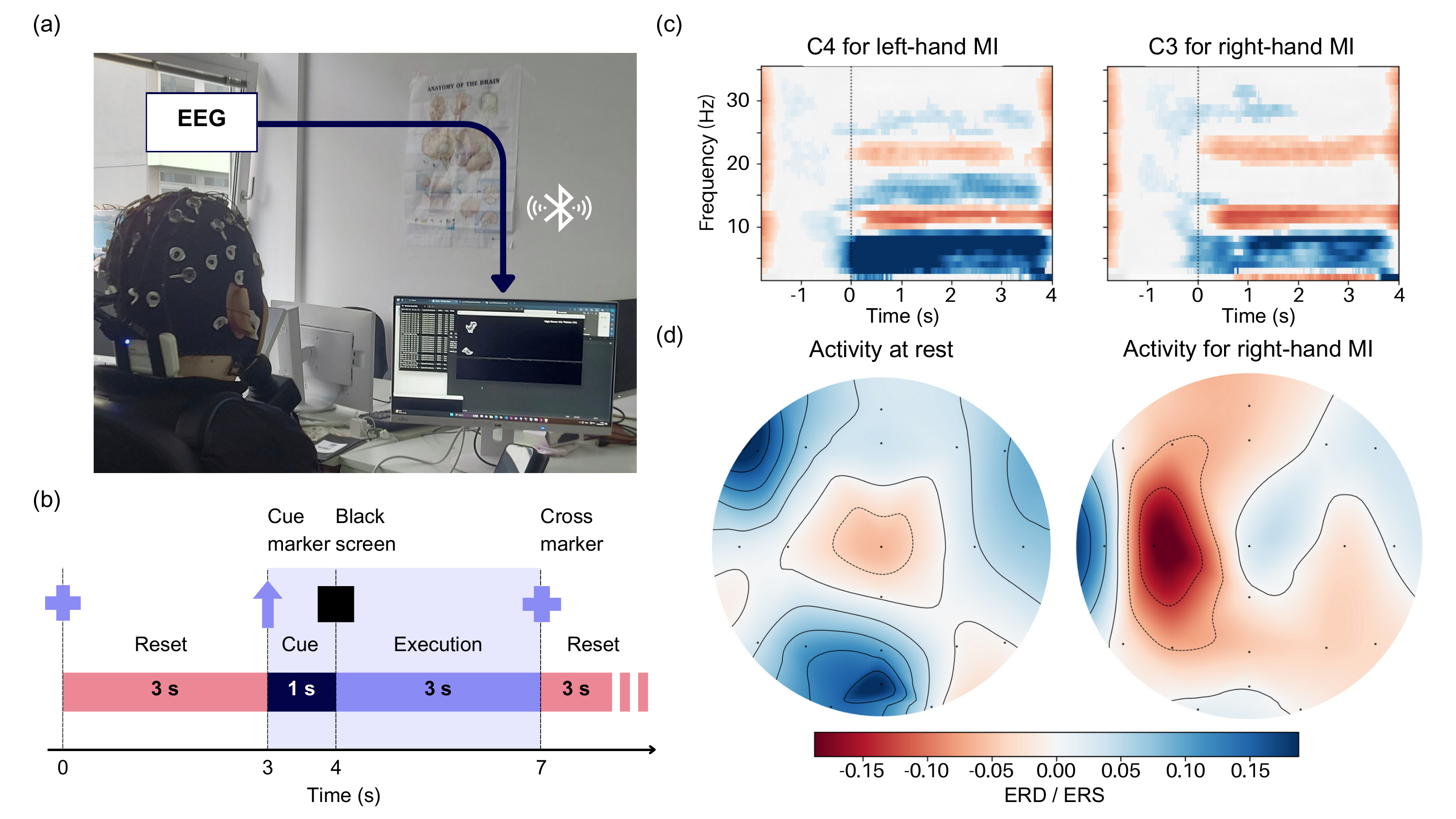} 
    \caption{(a) The pilot wearing the Smarting EEG from mBrainTrain, which collects data and wirelessly sends it to the computer via Bluetooth 2.1. (b) Time course for displaying cues to the pilot to collect data. First, the pilot sees a black screen with a cross marker. Then, a cue marker appears for one second. Depending on the cue, the pilot performs a different task. For example, a left arrow can denote left-hand motor imagery, a right arrow right-hand MI, and a circle rest. After the cue disappears, the pilot begins to perform the imagery or rest task for three seconds. Finally, once the cross marker reappears, they take a three-second break. (c) Event-related synchronisation (ERS) and desynchronization (ERD) over the course of an epoch during left-hand MI at the C4 electrode and right-hand MI at the C3 electrode. (d) Topoplot showing ERD/ERS activity over the entire brain before and during right-hand motor imagery. Activity is pronounced in the left hemisphere around the motor cortex.}
    \label{fig:fig1}
\end{figure}

For collecting data of different mental tasks, we employed an arrow cue paradigm (as shown in Figure \ref{fig:fig1}b). In response to cues displayed on the screen, presented as arrows pointing in one of the four cardinal directions, participants perform different mental imagery tasks.
The arrow cue paradigm is well-established in EEG-based BCI studies, indicating when to start and stop the mental imagery tasks \cite{Gwon2023_PublicMIDatasets}, \cite{Statthaler2017_CybathlonMirage91}.
The paradigm is executed as follows: 
First, the pilot sees a black screen with a cross marker. Then, a cue marker appears for one second. 
Depending on the cue, the pilot performs a different task. 
For example, a left arrow can denote left-hand motor imagery (MI), a right arrow right-hand MI, and a circle rest. 
After the cue disappears, the pilot begins to perform the imagery or rest task for three seconds. 
Finally, once the cross marker reappears, they take a three-second break.
\par
In addition to mental task data, we also collect calibration data for artefact removal, which entails the participant sitting as relaxed and calm as possible, avoiding movement and eye blinks for one minute \cite{Guarnieri2018}.
\par
Using this paradigm, we also tested a mental arithmetic task. 
For mental arithmetic, the pilot was asked to multiply four-digit numbers shown on a sheet of paper, in order to avoid the task relying heavily on memory. 
Combinations of mental imagery and mental arithmetic tasks were also evaluated, since different combinations can influence performance, both for the user and the BCI classification \cite{Kim2013}. 
\par
To ensure our pilots can perform mental imagery, we analyzed whether a task-corresponding signal could be found in the relevant frequency bands.
Motor imagery, or the mental rehearsal of movement, induces a localized reduction in power (event-related desynchronization) within the alpha (8-12 Hz) and beta (15-30 Hz) frequency bands \cite{Neuper2001_EventRelatedDynamics}. These bands can also be found in our pilots, as illustrated in Figure \ref{fig:fig1}c.
\par
Through initial testing with our pilot, we determined that leg and hand MI tasks produced the most consistent and distinguishable neural patterns, and the pilot expressed a clear preference for performing these tasks. A third “Rest” class was also included as a baseline condition.
As shown in Figure \ref{fig:fig1}c, event-related desynchronization and subsequent resynchronization are evident at electrode C3, with a clear reduction in alpha and beta power following the onset of MI. Furthermore, the topographical maps in Figure \ref{fig:fig1}d demonstrate that the spatial distribution of neural activity differs across tasks, reflecting task-specific cortical engagement. This shows that our pilot successfully generated neural activation patterns characteristic of motor imagery, as described in the literature \cite{Neuper2001_EventRelatedDynamics}.

\subsection{BCI System}
\begin{figure}
    \centering
    \includegraphics[width=1\linewidth]{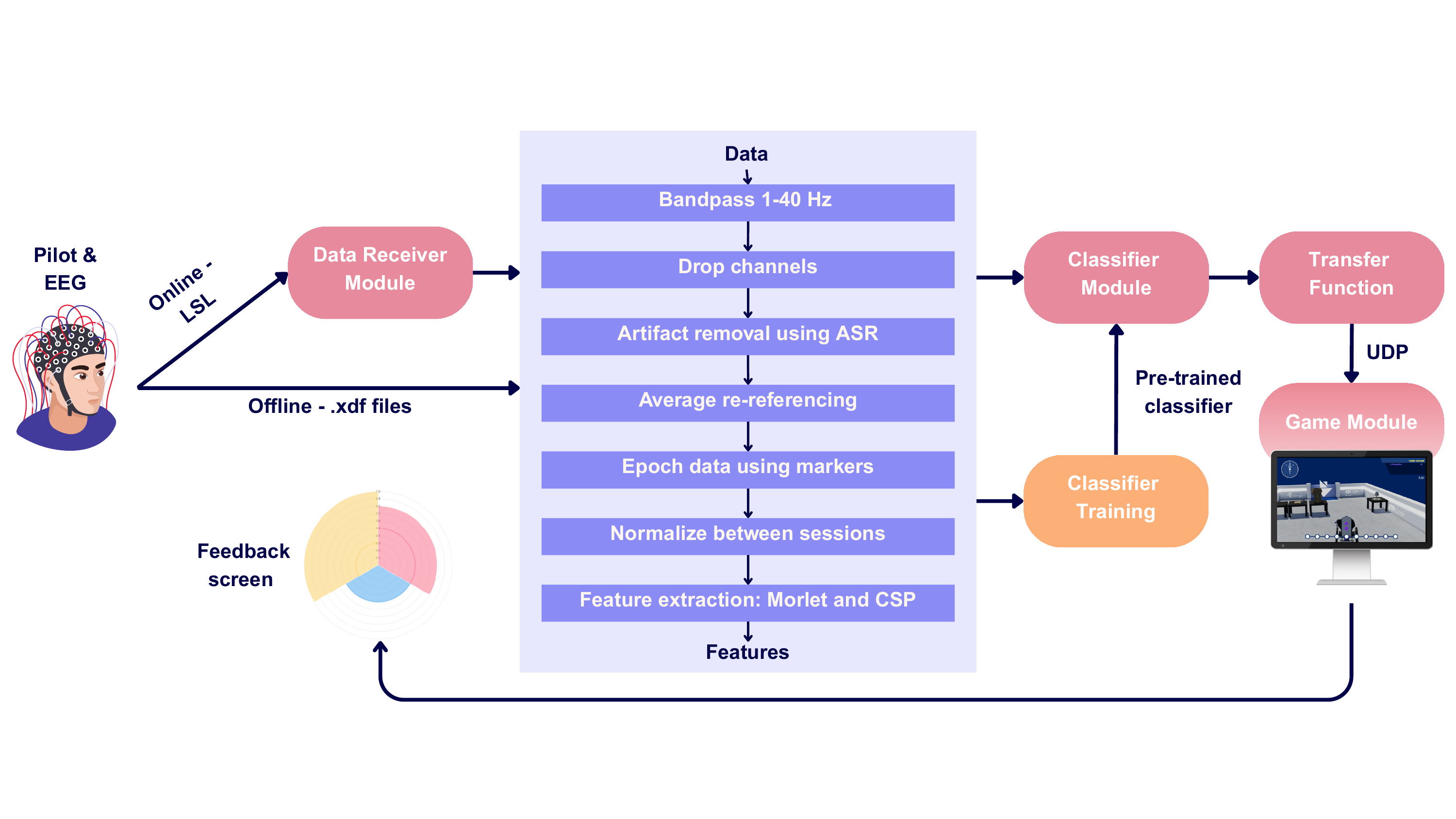}
    \caption{Overview of the brain-computer interface processing pipeline. The blue section illustrates the online and offline pipeline used for data processing, which includes EEG data preprocessing (bandpass filtering between 1-40 Hz, artifact removal via ASR, average re-referencing, and normalization across sessions), as well as feature extraction using Morlet wavelets and common spatial patterns (CSP). Classifier training (orange) is only performed in the offline pipeline. The red sections depict the online pipeline, where EEG signals acquired from the pilot are streamed via LSL, processed through the same preprocessing and feature extraction modules, and classified in real-time. The resulting predictions are transmitted via UDP to the feedback/game module, enabling closed-loop interaction.}
    \label{fig:fig2}
\end{figure}

Our BCI consists of a hardware component, the commercial 24-channel Smarting mobi EEG device from mBrainTrain, which transmits EEG signals and markers (task cues) to a computer via Bluetooth 2.1 (as shown in Figure \ref{fig:fig1}a), and a software component, our real-time pipeline (Figure \ref{fig:fig2}). Our pipeline supports two different modes: offline and online. In offline mode, we process previously collected labeled EEG data and train our classifier, which can then be used in the real-time (online) system. The central component of our pipeline is the asynchronous online processing, which receives the raw EEG signals and sends control signals to an effector device. Data preprocessing is the same in the offline and online modes of the pipeline. Both modes harness the preprocessing, feature extraction, and classifier modules described in the following sections.

\subsection{Preprocessing}
The pipeline begins by reading the raw EEG data files and corresponding event markers. 
It also loads one minute of calibration data and a mapping file that specifies which on-screen icons or marker labels correspond to the pilot's specific actions or motor imagery tasks. 
The raw EEG data are then band-pass filtered between 4 and 36 Hz using an infinite impulse response (IIR) Chebyshev type 2 filter \cite{Gramfort2021SelfSupervisionEEG}, \cite{Ouyang2025ProtocolEEG}, \cite{Delorme2004EEGLAB}.
The filter has an almost constant group delay of 500ms.
We forgo a notch filter since the bandpass filter sufficiently filters out the 50 Hz power line disturbance for our needs.
\par
After filtering, we crop the signals to reduce perturbations (e.g., boundary and adaptation effects) that may arise during filter initialization and transition periods \cite{Widmann2015DigitalFilterDesign}. 
The first 10 seconds of each recording are removed. 
Then, channel rejection is performed for channels exhibiting extreme artifacts (e.g., flatlines, excessive noise, or non-physiological spikes) to prevent distortion of the new reference estimate and to maintain the assumption of temporal independence underlying independent component analysis \cite{Delorme2004EEGLAB}.
\par
Artifacts such as eye blinks and muscle activity can significantly degrade the ability of a decoder to classify motor imagery \cite{Fatourechi2007EMGEOG}. 
To mitigate these effects, we applied Artifact Subspace Reconstruction (ASR) \cite{Mullen2013RealTimeModeling}, which was chosen due to its computational efficiency and suitability for real-time applications.
After ASR, the data are re-referenced using a common average reference (CAR) \cite{Ludwig2009CommonAverage}.
Re-referencing is performed at this stage because ASR requires data of full rank, and applying the reference afterward ensures that the rank is preserved during artifact removal \cite{asrpatent}. 
Following re-referencing, additional channels may be removed if they were initially required for reference computation but fail to meet the quality threshold for subsequent analyses. 
Subsequently, baseline correction and epoching are applied to segment the continuous data. When working with multiple datasets, the resulting epochs are further normalized to enable consistent comparisons across recordings.
\par
The preprocessing is kept the same across models and only differs for the EEGEncoder architecture presented below (see section \ref{sec:classifier}). 
Preprocessing parameters are presented in Table \ref{tab:preprocessing}.

\begin{table}[htbp]
    \centering
    \caption{Preprocessing parameters.}
    \begin{tabular}{c|c|c}
    \toprule
       Method & Parameter & Value \\
    \midrule
       Filter & Frequency range & 4-36 Hz \\
       Filter & Filter order & 16 \\
       ASR & Maximum dropout fraction & 0.4 \\
       ASR & Threshold & 3 standard deviations\\
       ASR & Method & Euclid \\
    \botrule
    \end{tabular}
    \label{tab:preprocessing}
\end{table}

\subsection{Feature Extraction}
During the development of our BCI pipeline, we integrated multiple feature extraction methods into our system. 
We use stacked Morlet \cite{Tallon-Baudry722} and Common Spatial Pattern (CSP) \cite{Koles1991QuantitativeExtraction}, \cite{MullerGerking1999OptimalSpatialFilters} features.
Morlet wavelets are a spectro-temporal feature. 
We calculate the single-trial power \cite{Gramfort2021SelfSupervisionEEG}, \cite{GramfortEtAl2013a}, using two cycles per frequency and only keeping every third output feature, discarding the rest.
CSP is a supervised method that requires labeled data and calculates spatial filters \cite{Blankertz2008OptimizingSpatialFilters}. 
We calculate the CSP features \cite{Gramfort2021SelfSupervisionEEG}, \cite{GramfortEtAl2013a}, saving the trained supervised decomposition for future use. 
The results are log-transformed and not regularized.
\par
We perform a stratified 80/20 train-test split for all models, maintaining prior data distributions. 
Finally, we apply windowing to extract more data points for training the classifier on.
\par
The feature extraction is performed the same way across all models, except the EEGEncoder, which is described below (see section \ref{sec:classifier}). 
Feature extraction parameters are presented in Table \ref{tab:featex}.

\begin{table}[htbp]
    \centering
    \caption{Feature extraction parameters.}
    \begin{tabular}{c|c|c}
    \toprule
        Method & Parameter & Value \\
    \midrule
       Morlet & Number of cycles & 2 \\
       Morlet & Decimation factor & 3 \\
       Morlet & Feature frequencies & 8-30 Hz \\
       CSP & Number of CSP components & 6 \\
    \botrule
    \end{tabular}
    \label{tab:featex}
\end{table}

\subsection{Classifiers}\label{sec:classifier}
Our BCI system supports the training and use of different machine learning and deep learning classifiers. 
We started with classical machine learning models. 
In particular, we utilized Support Vector Machines (SVM) extensively during the development process \cite{Pedregosa2011ScikitLearn}. 
SVM operates by finding a hyperplane that best separates different classes in a high-dimensional space, optimizing the margin between classes for robust classification results \cite{CortesVapnik1995SupportVectorNetworks}.
This makes it suitable for generalizing to small datasets, such as ours.
In addition to traditional approaches, we implemented a classifier consisting of three stacks of bidirectionally running S4D layers \cite{gu2022b} and a linear projection onto our set of classes. 
A state-space model, such as S4D, maps the history of an input sequence onto a set of basis polynomials \cite{gu2020}, \cite{gu2022trainhippostatespace}, which can be Legendre polynomials or sines and cosines \cite{gu2022b}, \cite{gu2022efficientlymodelinglongsequences}. 
The history map corresponds to a linear recurrence, enabling the hidden states to be updated online \cite{gu2020}. Updating the hidden state allows the model to retain information from previous time steps, which informs the classification of a new input. 
This approach is similar to classical signal processing techniques (e.g., the Fourier transform).
\par
The exact architecture is the following: first, the input features are mapped through a linear projection to \texttt{hidden\_dim} dimensions. This then passes through \texttt{num\_layers} of S4D blocks with skip connections and normalization. Afterwards, we perform temporal average pooling and perform logit classification for \texttt{num\_classes} to have a probability as an output.
Each S4D block consists of an S4D kernel that implements a diagonal state space model with matrices $A$ (diagonal with complex eigenvalues), $B$, $C$ (complex), and $D$ (skip connections), as well as the learnable discretization step size per channel, $dt$. The kernel is followed by a Gaussian error linear unit (GELU) activation function and a dropout layer with a \texttt{dropout\_rate}. The output projection of the S4D block is a Conv1D layer followed by a gated linear unit (GLU). We provide our parameters in Table \ref{tab:s4}. The implementation of the model is optimized through Vandermonde multiplication to compute the discrete convolution kernel and FFT-based convolution.

\begin{table}[h]
    \centering
    \caption{S4 classifier parameters.}
    \begin{tabular}{c|c}
        \toprule
        Parameter & Value \\ \midrule
        \texttt{num\_classes} & 3 \\
        \texttt{hidden\_dim} & 256 \\
        \texttt{num\_layers} & 3 \\
        \texttt{dropout\_rate} & 0.4 \\
        \bottomrule
    \end{tabular}
    \label{tab:s4}
\end{table}

\par
We incorporate Monte Carlo (MC) dropout to estimate the confidence of our model's predictions \cite{Gal2016DropoutBayesian}. 
By enabling dropout during inference, we generate multiple stochastic forward passes through the model, each time dropping out different neurons, approximating Bayesian inference. 
This results in a distribution of outputs from which we can calculate the mean prediction and its associated uncertainty. 
The confidence estimates provided by MC dropout are particularly valuable for identifying uncertain predictions, allowing for more informed decisions during both online and offline BCI tasks.
We save the classifier and can then use it in the real-time pipeline.
\par
To benchmark our motor imagery classification results, we reimplemented the EEGEncoder architecture proposed by Liao et al. \cite{Liao2025EEGEncoder} and trained it on our dataset, following the model configuration reported in the original publication. 
The EEG signals were first preprocessed by applying a 0.5-100 Hz band-pass filter and a 50 Hz notch filter; afterward, each channel was standardized to have zero mean and unit variance. 
The resulting signal was then segmented and fed into the model’s Downsampling Projector, an input module comprising three convolutional layers with average pooling and dropout to reduce the dimensionality and noise of the input signals. 
The projector’s output is subsequently passed to multiple parallel Dual Stream Temporal Spatial (DSTS) blocks. 
These blocks combine a temporal convolutional network (TCN) path with a stable Transformer path that employs pre-normalization via Root Mean Square Layer Normalization (RMSNorm) and uses Swish-Gated Linear Unit (SwiGLU) activations. 
In line with the authors’ reference architecture, we used five parallel DSTS branches, four Transformer layers, and two attention heads.
\par
For further comparison, we employed EEGNet, a compact convolutional neural architecture specifically designed for EEG-based BCI applications \cite{Lawhern2018EEGNet}. 
EEGNet applies temporal convolutions to learn frequency-specific filters, followed by depthwise and separable convolutions that extract spatial patterns while remaining highly parameter-efficient. 
Batch normalization and average pooling complement the convolutional blocks, after which a fully connected layer carries out the classification. 
Compared to more recent deep-learning approaches, EEGNet is often outperformed on large-scale EEG datasets \cite{Liao2025EEGEncoder}, \cite{zhao2024ctnet}. 
Nevertheless, it remains a widely used and well-validated baseline due to its architectural simplicity, its suitability for small-sample EEG scenarios, and its extensive evaluation across multiple BCI benchmarks. 
In our study, the model was trained directly on normalized raw EEG windows extracted from the participant’s recording. 
No additional preprocessing or feature extraction was applied, as EEGNet is designed to operate on raw spatio-temporal EEG tensors. 
Training was performed using the Adam optimizer with an initial decaying learning rate of $2\times 10^{-4}$ and a minibatch size of 64, with dropout (0.2) used for regularization as well as early stopping based on validation performance. The best model was obtained after 50 epochs.

\subsection{Online/Real-time pipeline}
Our real-time system consists of separate modules that run in parallel (Figure \ref{fig:fig2}). 
The first is the data receiving module. The Smarting EEG software uses the Lab Streaming Layer (LSL) ecosystem to announce this data stream to machines in our network \cite{10.1162/IMAG.a.136}. 
Our first module receives this announcement and subscribes to the stream. As the Smarting software receives data from the EEG device, it creates chunks of data and publishes them for our module to receive. 
This module was developed using PyLSL\footnote{See https://github.com/labstreaminglayer/pylsl (last accessed November 12, 2025)} to receive our data, buffer it, and propagate it down our pipeline to the preprocessing module, which utilizes ZeroMQ\footnote{See https://zeromq.org (last accessed November 12, 2025)}, a brokerless asynchronous messaging library for inter-process communication within our pipeline. 
The preprocessing module, which receives the chunk of data, applies the same processing steps as in the offline pipeline and then publishes them for the next module to receive. 
The next module contains the classifier trained using the offline pipeline. 
The classification output of the model on the preprocessed and feature-extracted data snippet is sent to the transfer function. 
\par
The transfer function maps the three-class output of the classifier onto up to 5 control signals. 
Here, we use only three control signals. 
The transfer function consists of a rolling buffer that fills up with the probabilities of classification obtained by taking the average over the classification results given by the model. 
The most probable class is chosen as an output to the model. 
This essentially constitutes a downsampling and smoothing operation, since we have a high classification rate, which would be unusable in real-world experiments.

\subsubsection{Games}
\begin{figure}
    \centering
    \includegraphics[width=1\linewidth]{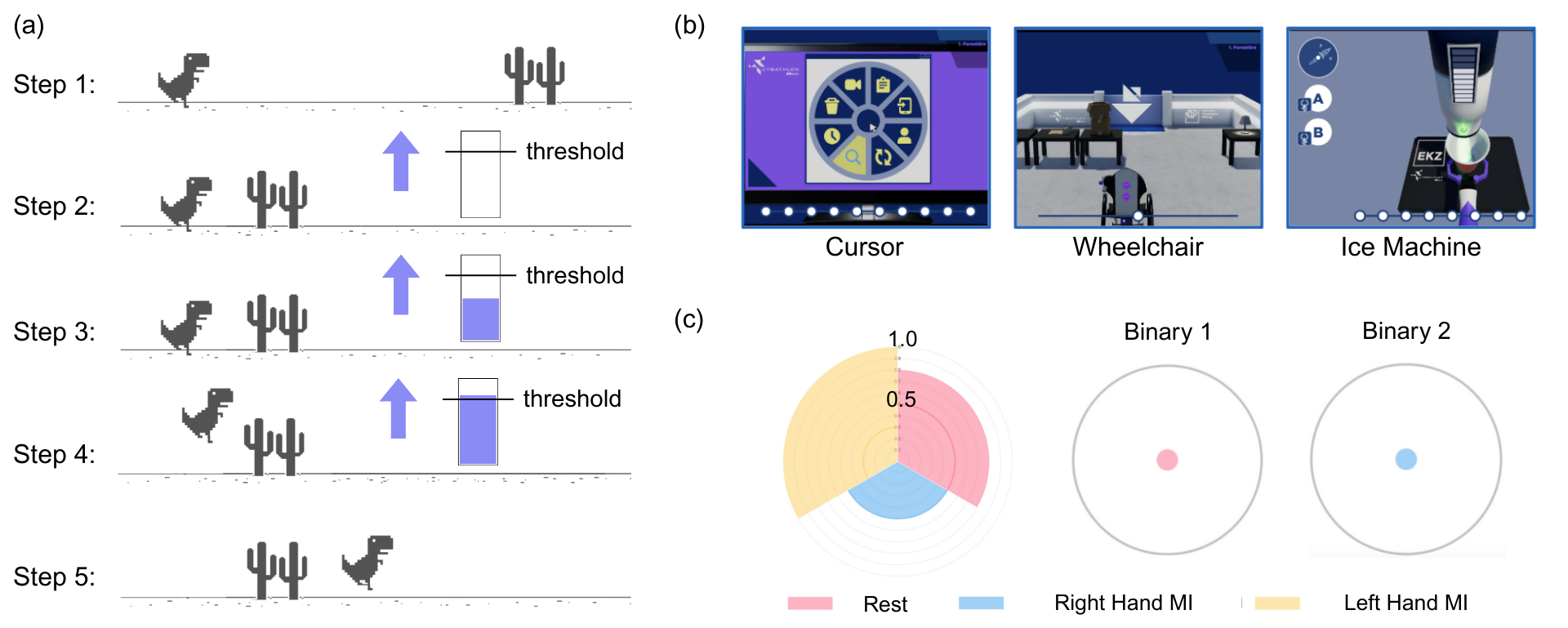}
    \caption{(a) This part of the figure shows one progression of the Dino game. In step 1, the dinosaur is running. When it reaches an object it needs to jump over, such as the cactus, it stops, and a task is indicated via the arrow marker (step 2). When the classifier classifies that the pilot is producing the imagery denoted by the marker, the bar fills up (step 3). If the threshold is reached, the dino jumps over the object (step 4). Finally, the dino keeps running until it reaches the next object (step 5). If the threshold is not reached, then the dino does not jump and loses one health point. (b) Three of the possible sub-games/tasks of the cybathlon game as played in the competition. In the Cursor game, the pilot needs to move the cursor to the highlighted block on the screen and use one of the binary commands to click on it. For the Wheelchair task, the pilot navigates the wheelchair through the room to the exit marked with an arrow. In the Ice Machine task, a robotic arm with a cup is placed under the sensor of an ice machine to collect a set amount of ice. Misclassification leads to the cup tipping and ice falling out. (c) The feedback screen visualises the classification output and thresholds of the transfer function to the pilot. On the left, the large circle shows the different classes being classified via the colourful wedges. If one of them reaches over the threshold in their third, then the corresponding control signal is sent to the game. The two circles on the right fill up based on the class that is classified, and once the circle is full, one of the binary signals is sent to the game.}
    \label{fig:fig3}
\end{figure}

The first game, shown in Figure \ref{fig:fig3}a, is an adapted version of the Chrome Dino game, developed as part of a human-centered approach to BCI training. The design aimed to strike a balance between experimental control and user engagement, enabling the pilot to remain actively involved in the training process while allowing for systematic evaluation of decoding performance.
\par
The game features a QuickTime mode, in which the character pauses or slows down before an obstacle to create a temporally controlled interaction window. 
During this three-second phase, a cue marker appears, prompting the pilot to perform a specific motor imagery task. 
\par
The classifier output leads to an increase in the height of a bar visible on screen if the classification matches the cue marker. When misclassifications occur, a visual control bar decays in height, providing immediate and intuitive feedback on classifier performance. If the bar reaches a certain threshold within a predefined time (e.g., 3 seconds), then the character jumps over the obstacle. If the bar is below the threshold, no jump occurs. A jump is recorded as a successful task completion in online testing of the classifier, and no jump is consequently a failed task in the online assessment of the classifier. 
\par
This interactive paradigm reflects key human-centered design principles by integrating real-time feedback, self-pacing, and user participation in the control loop. 
It allows the pilot to experiment with different motor imagery strategies, understand the system’s responses, and develop a sense of agency and control, factors known to enhance sustained engagement. 
At the same time, the paradigm retains the analytical rigor of cue-based tasks, as all events are time-stamped and synchronized for detailed offline evaluation. 
The game thus serves as both a training tool and an experimental platform for refining motor imagery decoding before deployment in the Cybathlon task.
\par
The second game, shown in Figure \ref{fig:fig3}, was developed for the 2024 Cybathlon BCI race and consists of five distinct tasks, each repeated twice, for a total of ten tasks to be completed within an eight-minute time frame. 
The tasks include navigating a wheelchair through a room, maneuvering the wheelchair while avoiding a moving vacuum cleaner, performing a computer screen-based task, operating a key lock, and collecting ice from an ice machine. 
Examples of these tasks are illustrated in Figure \ref{fig:fig3}b.
\par
Control across all tasks is achieved using four input signals: two continuous signals that govern movement along the X-axis and Y-axis (both from $-1$ to $1$), and two binary signals (A and B). 
The continuous Y-axis input drives forward or backward motion of the avatar (wheelchair, cursor, or robot arm), while the X-axis input rotates the avatar's orientation clockwise or counterclockwise. 
The binary inputs are used to activate task-specific buttons or functions depending on the context of each task.

\subsubsection{Feedback screen}
The feedback application was implemented as a web-based interface running alongside the main BCI pipeline. 
It receives the output of the transfer function -- the same control signal sent to the Cybathlon game -- along with relevant metadata, such as the most recent classification label. 
The application, developed using the Flask framework, can dynamically render different visualization templates (e.g., bar or pie charts) based on the incoming data. 
Data transmission occurs via a secure local WebSocket connection to ensure low-latency communication within the local network.
Providing users with real-time feedback has been shown to play a crucial role in improving BCI control performance and facilitating motor imagery learning \cite{Neuper2001_EventRelatedDynamics}, \cite{Jeunet2016MiBCI}, \cite{Lotte2013FlawsBCITraining}. 
Feedback enables users to understand the system's responses, adapt their strategies, and reinforce successful mental states through operant conditioning. 
Moreover, a feedback interface can be used to calibrate the threshold parameters of the transfer function before or during real-time operation, ensuring more robust and individualized control.
During the Cybathlon competition, however, the feedback screen was not used. 
The competition setup required the pilot to focus entirely on the official game display, and additional feedback could have introduced distraction or cognitive overload under high-pressure conditions. 
Nevertheless, in training and development settings, the feedback interface remains a valuable tool for performance optimization and calibration.

\section{Results}\label{results}
\subsection{Offline Classification}
\begin{figure}
    \centering
    \includegraphics[width=0.95\linewidth]{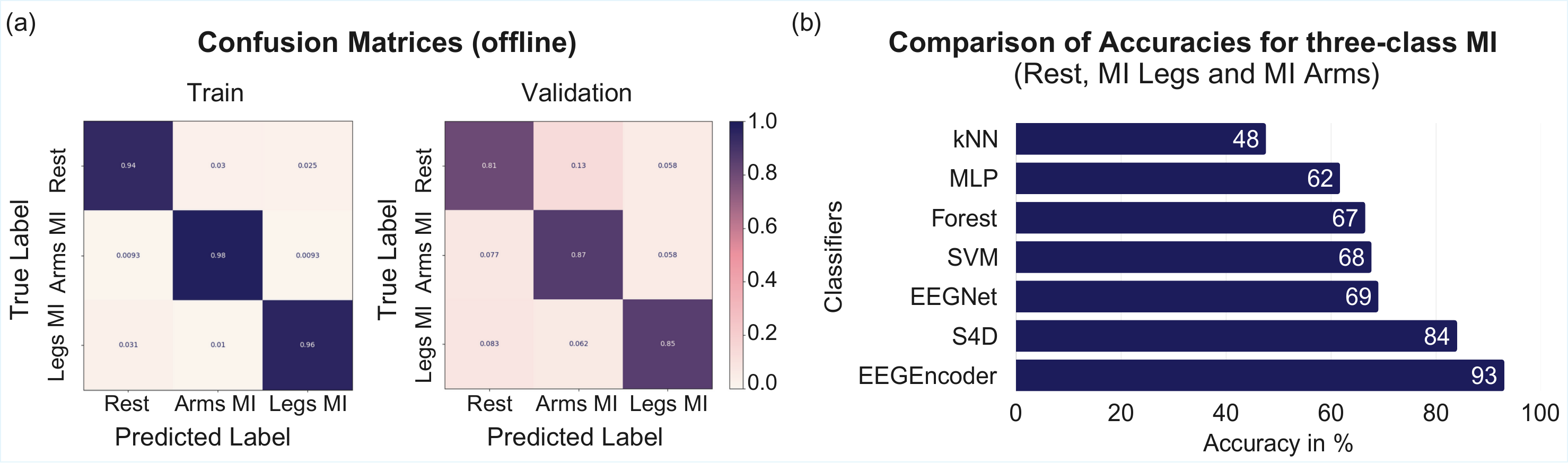 }
    \caption{(a) On the left is the confusion matrix for the S4D-layer-based model on the training data, and on the right is the confusion matrix for the testing/validation of the data. The 190 cued samples used here correspond to the data obtained during the two Cybathlon competition days in five sessions. (b) Comparison of accuracies for three-class MI. The S4D-layer-based model and EEGEncoder outperform all other baselines. The data is the same as in (a).}
    \label{fig:fig4}
\end{figure}

Training the S4D model on data from our Cybathlon pilot yielded an accuracy of 84\% for three-class motor imagery classification (Figure \ref{fig:fig4}a). 
The data used for training correspond to the data used for training the models during the two Cybathlon competition days, in five separate sessions. 
This data comprises 190 cued samples, which corresponds to 760 samples input into the model, 80\% of which are used for training.
\par
To contextualize this performance, we compared it with several machine learning baselines (SVM, random forest, AdaBoost, k-NN, MLP) as well as EEGNet and the recently proposed Transformer-based architecture tailored for motor imagery, EEGEncoder \cite{Liao2025EEGEncoder} (Figure \ref{fig:fig4}b).
The machine learning-based models underperformed the deep learning-based classifiers, achieving accuracies from 48\% for the kNN and up to 68\% for the SVM model.
The EEGNet model achieved 69\% accuracy, using the same preprocessing parameters as the S4D-layer-based model.
The EEGEncoder architecture achieved an offline accuracy of 93\% on this pilot's data, outperforming the S4D-layer-based model. 
\par
We further validated the performance of the decoders on data from an additional pilot, where the EEGEncoder and the S4D-layer-based model obtained 72\% and 71\% accuracy, respectively.
The data from this pilot were obtained on three consecutive days.
The data consists of 720 cued samples, corresponding to 2880 samples that are input into the model, 80\% of which are used for training.

\subsection{Real-time performance}
\begin{figure}
    \centering
    \includegraphics[width=0.85\linewidth]{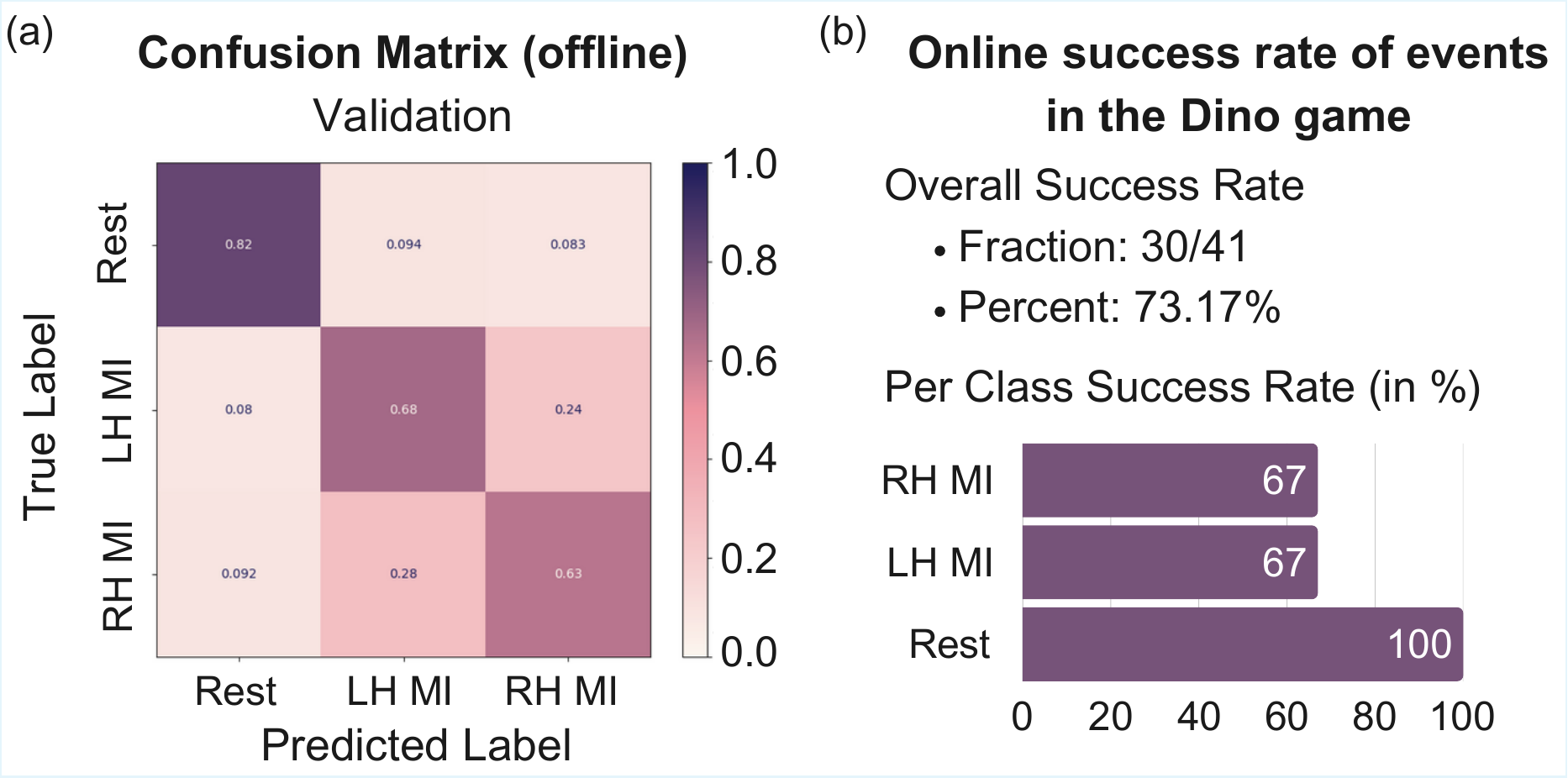}
    \caption{(a) Testing confusion matrix of the offline classifier training for 720 samples from pilot 2. The data was obtained on three consecutive days. The total accuracy is 71\%. (b) Online success rates from pilot 2 playing the Dino Game. The pilot achieves 73.17\% accuracy, matching the results from offline training performed on the same day.}
    \label{fig:fig5}
\end{figure}

As an assistive technology, BCIs are only useful if they function in real-time. 
To determine and quantify the effectiveness of our S4D-based decoder in real-time, we examine the success rate of jumps in the Dino game, as well as its online accuracy. 
We use the Dino game because it provides us with markers/labels for the mental task the pilot was performing during a given time. 
In the preliminary testing (see Figure \ref{fig:fig5}a), we gathered 42 trials while a pilot was playing the Dino game. 
Thirty-one of these were successful, specifically 100 percent of the rest tasks, and 67\% of the right and left hand motor imagery tasks were successful. 
When compared with the offline accuracy of 71\%, we were able to translate this into a very similar success rate of 73\% online.
While the Cybathlon game successfully emulates real-world scenarios, it does not lend itself well to detailed online evaluation of the classifier since we do not have a ground truth of what the person was thinking to compare the classification output to. 
With the Cybathlon game, we can only examine the number of possible subtasks that were successfully completed. 
In the competition itself, our pilot was able to solve one task in 8 minutes. 
With more training, playing the game, and utilizing additional data to train our classifiers, he is now able to solve 3 subtasks in 8 minutes. 

\subsection{Latency Analysis}
\begin{figure}
    \centering
    \includegraphics[width=1\linewidth]{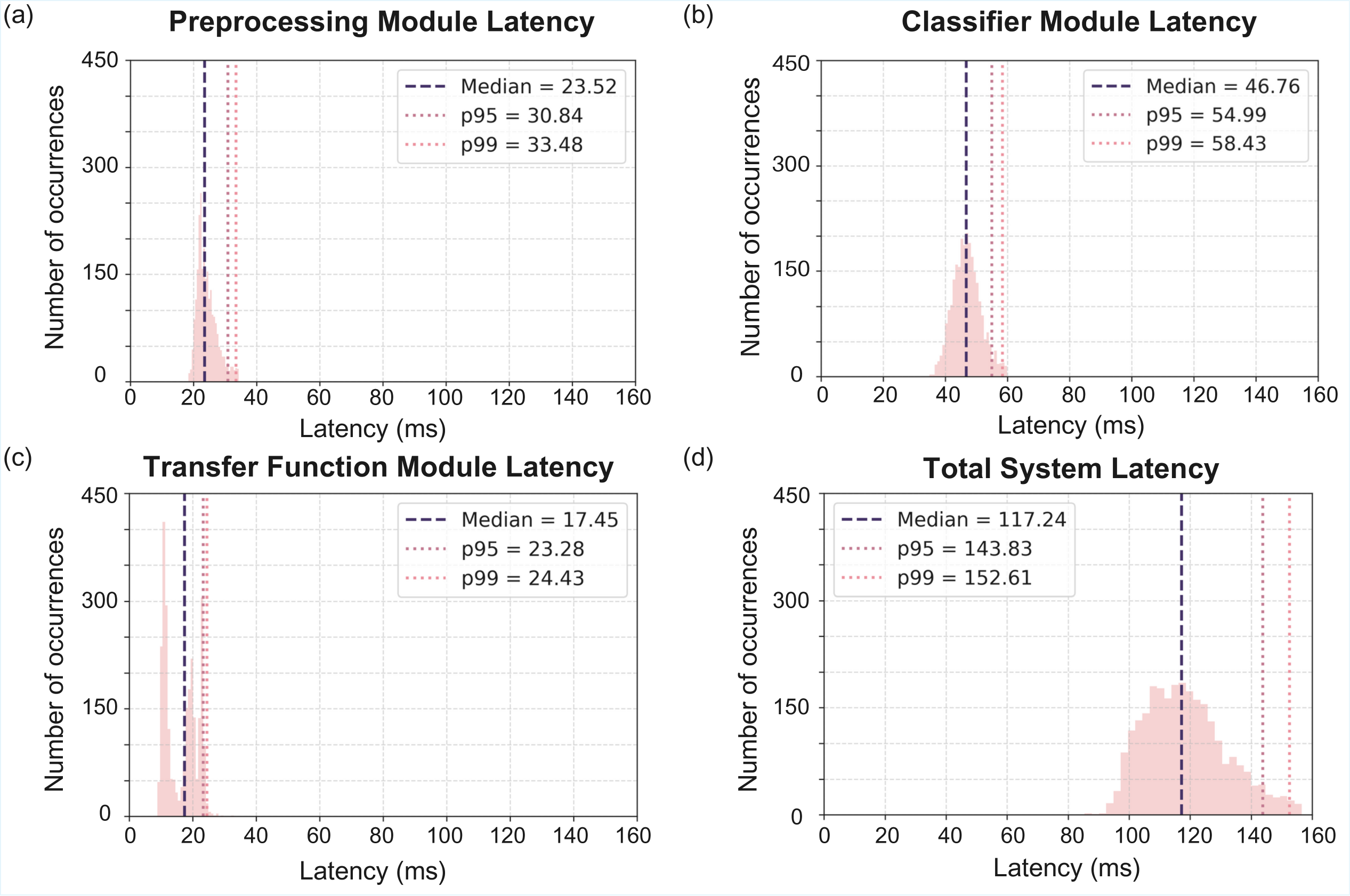}
    \caption{Latency distributions for all modules of the BCI pipeline. Each subplot shows the median (blue dashed line) and the 95th and 99th percentiles (red dotted lines) of the latency distribution for (a) preprocessing, (b) classifier, (c) transfer function, and (d) total system latency.}
    \label{fig:fig6}
\end{figure}

To assess the real-time performance of the system, we implemented an end-to-end computational latency monitoring pipeline encompassing all modules of the BCI. 
Latency timestamps were recorded at four key processing stages: (1) EEG acquisition, (2) preprocessing, (3) classification, and (4) transfer function. 
These timestamps were collected continuously during online operation, and the resulting timing data were analyzed offline to characterize inter-module delays. 
Figure \ref{fig:fig6} summarizes the computational latency distributions across consecutive modules. 
The median delay at the preprocessing stage was $23.52$ ms (p95 = $30.84$ ms), at the classification stage $46.76$ ms (p95 = $54.99$ ms), and at the transfer function $17.45$ ms (p95 = $23.28$ ms). 
The total end-to-end computational latency, measured from EEG acquisition to the transfer function output, was $117.24$ ms (p95 = $143.83$ ms).
\par
Additional methodological latency stems from the group delay of the filter, which is $500$ ms, and from the windowing procedure, which introduces a further $1000$ ms. 
This results in a total delay of $1617$ ms, or a theoretical information transfer rate (ITR) \cite{Billinger2013, SADEGHI2019101607} of $17.57$ bits per minute. 
We call this the theoretical ITR, since for online gameplay, more than one classification is needed. 
A more realistic estimate should take into account the multiple classifications needed. 
From experimental observations, this can range from $1.5$--$3$ seconds, resulting in an experimental ITR of $9.11$--$6.15$ bits per minute.

\section{Discussion}\label{discussion}
The primary design goal of our system was to maximize its versatility and applicability across different use cases. 
To this end, we adopted a task-agnostic BCI paradigm, i.e., the performed task can be chosen flexibly by the user, depending on their needs. 
This increases the system's usability, as different users may not be able to produce a classifiable activation for the same task. 
While this choice inherently increases the complexity of classification compared to evoked potential BCIs, such as SSVEP, which typically achieve approximately 5-15\% higher accuracy on average \cite{Guger2012HowManySSVEP}, \cite{Liu2022ReviewSSVEP}, we considered it a more practical approach for future real-world applications, where reliance on visual stimuli or a display screen would be impractical.
\par
Our two main test pilots attempted various mental tasks: MI for the left and right hands, MI for the left and right halves of the body, MI for the arms and legs, and mental multiplication. 
While one pilot was able to produce consistent MI for the legs, the second pilot was not able to do so. 
Hence, one chose the left-vs-right body half MI, including the legs, and the second only performed the hand MI.
During the human-centered task design approach, mental multiplication was quickly abandoned because multiplying large numbers required significantly more time (up to ten seconds) than the three seconds chosen for the online EEG BCI paradigm, making it uncomfortable for the pilots. 
On the other hand, smaller numbers would have transformed the task into a purely memory-based retrieval. 
This demonstrates the importance of validating the pilot's ability to perform specific mental tasks to produce activation, as well as the importance of tailoring BCIs to user capabilities.
\par
The flexible structure of our pipeline allows the pilots to give feedback, which can immediately be integrated into the system. 
Therefore, the experimental sessions also consisted of regular discussions about the personal preferences of the pilots and the technical possibilities we have. 
This is what we believe to be human-centered design: One cannot expect a device to be used if the users are not directly involved in the production. 
In our case, this manifested through pilots implementing features in the games or in the pipeline themselves and testing them in their next experiment. 
Through this iterative and collaborative approach, we successfully integrated the diverse perspectives of patients and healthy participants into our pipeline.
\par
We acknowledge that the number of participants (N = 2) is relatively small. 
Nevertheless, we believe that our system can be expanded to a wider user base, and we will strive to do so for future work. 
First, since our pipeline is, in principle, task agnostic, a pilot does not necessarily need to be able to produce motor imagery, as other tasks could be devised. 
Second, the modularity as well as the easy and fast parametrization enable us to quickly react to different user performances. 
A practical example: if certain classes are overclassified during online gameplay, the thresholds for how many classifications are needed to make the dino jump can be updated during the gameplay, without the need to restart. 
A current limitation we face is the need to retrain the classifier for every gameplay session. 
This results in an overhead of approximately 30 minutes before the system can be used online. 
This could deter new users from using the system. 
Combined with the need for a few sessions to get accustomed to the system and learn how to perform motor imagery effectively, we believe this might limit the interested user base. 
To mitigate this factor and improve our system, we are considering incorporating transfer learning methods based on Riemannian geometry in the future. 
This would enable us to train one large classifier that can be retrained for specific users with a smaller sample of data. 
This would also help for repeated sessions for a single user, as the same principle could be applied \cite{riemanntransferYing}, reducing the impact of possible data shifts \cite{RAZA2019154}. 
\par
One main bottleneck we face with our system is delays. 
This reduces the usability of our system for daily use. 
Although the computational delay is comparatively low, the methodological delay is a significant issue.
To have a BCI that can be used in closed-loop systems, delays below $250$ ms are recommended \cite{sahar2016, xu2014, xu2016}.
To reduce the delay, one could consider using deep learning frameworks that implicitly learn the filtering as one of their layers. 
This would remove the $500$ ms group delay from the Chebyshev filter.
In principle, our S4D model could already perform this, since it can be interpreted as a convolution \cite{gu2022efficientlymodelinglongsequences}, \cite{STUCHI2024111443}. 
This could help in cutting overall delay time.
This would also remove any nonlinearities in the group delay, which might negatively affect the timeseries aspect of the S4D-layer-based model through frequencies being assigned to the wrong point in time.
However, since the used filter exhibited a relatively constant group delay, this was not an issue in our case.
\par
The delays also contribute to the comparatively low Cybathlon performance: simpler tasks, such as the moving wheelchair task, could be solved, as there was no constant movement.
However, with more complex tasks, which require more precision, performance rapidly dropped, prompting us to skip these in the interest of time. 
Therefore, our system can already complete time-insensitive tasks, but struggles with more time-sensitive ones.
This also highlights a limitation of testing BCIs in software: while games provide much leeway to ignore smaller errors, hardware applications mostly do not, without severely impacting user safety. 
For future work, we will thus strive to reduce latency to a minimum.
Aside from latency, other factors, such as accuracy or false classification of other classes, can also reduce the effective control over the system. Improving those while maintaining the same latency would thus also improve overall system control.
\par
The necessity of feature extraction methods was tested in an ablation study. 
We found that using only Morlet wavelets or CSP achieved respective accuracies of 82.9\% and 66.6\% on the dataset of the patient. 
This compares to an accuracy of 84.1\% for the combined features. 
This improvement is relatively small, suggesting that CSP features might not be necessary.
Future work will explore feature selection methods, such as the Fisher score \cite{fisherscore}, to further improve the number of features used.
\par
The S4D-layer-based model, which has not been used for motor imagery classification before, consistently achieves high accuracies and has demonstrated success in real-time use. 
This is accompanied by short training times of around 5 to 15 minutes on readily available high-performance laptops, depending on whether CUDA support is available. 
This is important, since training large models over multiple hours is not easily feasible, due to experimental time constraints and possible distribution shifts \cite{RAZA2019154}. 
While the EEGEncoder achieved higher accuracies in offline testing, the longer training times (up to two hours) could be a mitigating factor, rendering it unusable for time-constrained experiments.
Given the strong offline performance of the EEGEncoder architecture, future work should focus on assessing its real-time performance and integration into our BCI system. 
This could also include trading a few percentage points in accuracy for shorter training times. 
The machine learning models we tested required relatively short training periods (2-5 minutes) and are simple to implement, without the need for performant hardware. 
The EEGNet model falls somewhere in between the machine learning models and the S4D-layer-based model in terms of training times.
\par
Recent MI-BCI studies employing transformer-based architectures have reported strong performance on standard benchmark datasets such as BCI Competition IV-2a and IV-2b. For example, MSCFormer achieved average accuracies of 82.60\% on the four-class IV-2a dataset and 88.00\% on the two-class IV-2b dataset \cite{msconv}, while CTNet reported subject-specific accuracies of 82.52\% and 88.49\% on IV-2a and IV-2b, respectively. In comparison, our approach achieves an average accuracy of 84\% on a three-class motor imagery dataset \cite{zhao2024ctnet}. Although these results are not directly comparable due to differences in class cardinality, dataset characteristics, and evaluation protocols, the obtained performance places our method within the range of recently reported state-of-the-art MI-BCI decoding approaches.
\par
Based on this discussion, choosing a model becomes a trade-off between accuracy, training time, and feasibility to run the model on available hardware in online settings.
This highlights the importance of going beyond offline accuracies to real-world, real-time validation. 
For our system, the S4D-layer-based model performed well, as its accuracy fell in line or outperformed other systems in both offline \cite{Lawhern2018EEGNet, Tabar2016DeepLearningEEG, 10.1371/journal.pone.0268880, 10.3389/fninf.2022.961089} and online \cite{10.3389/fninf.2022.961089, signals4010004, JAFARIFARMAND2020101749} classification settings. The training times did not negatively affect experiment execution, and we deem our hardware sufficient and accessible, making it a good option for our use case.

\section{Conclusion}\label{conclusion}
In this work, we presented a modular, online EEG-based BCI framework designed to investigate the portability, usability, and adaptability of our setup. 
Our system was developed in the context of the Cybathlon 2024 competition, which provided both a technically challenging and user-driven environment to test the practical feasibility of our system.  
By integrating a flexible architecture that allows for independent optimization of acquisition, preprocessing, classification, and control mapping modules, we demonstrated that high performance and low-latency online operation can be achieved on affordable, mobile-compatible hardware.
\par
Our system exhibits low computational latency, with further improvements possible through adapted filtering and windowing procedures. 
We implement a deep learning classifier, the structured state-space sequence, which has not been previously used for EEG motor imagery classification. 
We achieve sufficient classification performance while maintaining short training times. 
We also test an EEGEncoder architecture for EEG data published after the Cybathlon. 
While it outperformed the S4D-layer-based model, it also faced higher training times, paving the way for future optimization.
\par
Beyond technical performance, a major contribution of this work lies in its human-centred, iterative design approach. Close collaboration with our tetraplegic pilot informed the design of the training interface and the development of real-time feedback mechanisms. The development of a training game provided continuous engagement, promoting user learning, system calibration, and feedback-driven adaptation. This emphasis on usability and user experience directly addresses one of the main limitations of current BCI research, the gap between high laboratory accuracy and limited real-world applicability.
\par
Our findings underscore the importance of modularity and adaptability in BCI systems. By designing each module to function independently and in parallel, we make it easier and more feasible to substitute or upgrade components, such as preprocessing algorithms, classifiers, or feedback modalities, without compromising the overall system integrity. Furthermore, the system's modularity could provide a foundation for more scalable BCI solutions that can be personalized for different users, signal modalities, or assistive technologies.
\par
Future work will focus on improving robustness and user adaptation through hybrid feature integration (e.g., combining EEG with peripheral physiological signals), adaptive learning techniques, and personalized calibration routines that dynamically adjust to individual signal variability. Additionally, we aim to extend the system’s capabilities beyond competitive environments to daily assistive scenarios.

\backmatter

\section*{List of Abbreviations}
\begin{itemize}
\item BCI: brain-computer interface
\item EEG: electroencephalography
\item MI: motor imagery
\item S4D: diagonalized structured state-space sequence
\item ASR: artefact subspace reconstruction
\item DSTS: dual stream temporal spatial
\item LSL: lab streaming layer
\item IIR: infinite impulse response
\item FIR: finite impulse response
\item GLU: gated linear unit
\item GELU: Gaussian error linear unit
\item TCN: temporal convolutional network
\item CSP: common spatial patterns
\item SVM: support vector machine
\item MC: Monte Carlo
\item ITR: information transfer rate

\end{itemize}
\section*{Ethics approval and consent to participate}
This work reports on the self-chosen participation of a pilot in a championship; hence, ethical approval was not applicable. All participants gave their consent to take part in the experiments. 

\section*{Consent for publication}
Consent for publication was obtained from all participants.

\section*{Availability of data and materials}
The datasets created for and subsequently used and analyzed during the current study are available under \cite{neurotum_e_v_2025_18087806}. 
The pilot IDs are P1 for the patient and P2 for the healthy participant.
The code used for this study will be made available upon reasonable request.

\section*{Competing interests}
The authors declare that they have no competing interests.

\section*{Funding}
Funding to participate in the Cybathlon was obtained from Freunde der TUM e.V. and the Cybathlon. The used EEG device was donated to neuroTUM e.V. by mBrainTrain, Belgrade, Serbia.

\section*{Author contribution}
IWT analyzed the data, implemented the Dino Game, was a major contributor in writing the manuscript, and led the organization. 
NCT worked on the signal processing, validated the system as a pilot, and was a major contributor in writing the manuscript. 
AMF designed the tasks and was a major contributor in writing the manuscript.
IC wrote a substantial part of the pipeline and led the organization.
LJ validated the system as a pilot and contributed to the design of the experimental tasks.
SH designed the experimental tasks.
TEH implemented the S4D model and edited the manuscript.
DP worked on software engineering 
MM implemented the feedback screen.
PM implemented the EEGEncoder model and wrote the corresponding sections of the manuscript.
GVK implemented the EEGNet model and contributed the corresponding section in the manuscript.
OSR performed latency analysis.
TL implemented the transfer function from the classifier to the different games.
FP contributed to the experimental design.
EAF implemented the LSL communication.

\section*{Acknowledgements}
We thank all past and present members of the neuroTUM e.V. BCI Team for their contributions. We would also like to thank the Chair of Neuroelectronics and the Institute of Cognitive Systems at the Technical University of Munich for their support. We thank Paolo Favero and Delfina L. Taskin-Espinoza for offering their guidance to the club during the writing process. 

\section*{Authors' information}
neuroTUM e.V. is a student-led non-profit research club/organisation working on projects at the intersection of neuroscience and engineering. We are based in Munich and were founded in 2022.

\bibliography{sn-bibliography}

@inproceedings{sahar2016,
    author = {Selim, Sahar and Tantawi, Manal and Shedeed, Howida and Badr, Amr},
    booktitle= {Proceedings of the International Conference on Advanced Intelligent Systems and Informatics 2016},
    year = {2016},
    month = {10},
    pages = {555--565},
    title = {{Reducing execution time for real-time motor imagery based BCI systems}},
    doi = {10.1007/978-3-319-48308-5_53},
}

@article{xu2014,
    author = {Xu, Ren and Jiang, Ning and Lin, Chuang and Mrachacz-Kersting, Natalie and Dremstrup, Kim and Farina, Dario},
    year = {2014},
    month = {02},
    pages = {288--296},
    title = {Enhanced Low-Latency Detection of Motor Intention From {EEG} for Closed-Loop Brain-Computer Interface Applications},
    volume = {61},
    journal = {IEEE Transactions on Bio-Medical Engineering},
    doi = {10.1109/TBME.2013.2294203}
}

@ARTICLE{xu2016,
    AUTHOR={Xu, Ren  and Jiang, Ning  and Mrachacz-Kersting, Natalie  and Dremstrup, Kim  and Farina, Dario },
    TITLE={Factors of Influence on the Performance of a Short-Latency Non-Invasive Brain Switch: Evidence in Healthy Individuals and Implication for Motor Function Rehabilitation},
    JOURNAL={Frontiers in Neuroscience},
    VOLUME={Volume 9 - 2015},
    YEAR={2016},
    DOI={10.3389/fnins.2015.00527},
}

@article{Jaeger2023,
  author       = {Lukas Jaeger and Roberto de Souza Baptista and Chiara Basla and Patricia Capsi-Morales and Yong Kuk Kim and Shuro Nakajima and Cristina Piazza and Michael Sommerhalder and Luca Tonin and Giacomo Valle and Robert Riener and Roland Sigrist},
  title        = {How the CYBATHLON Competition Has Advanced Assistive Technologies},
  journal      = {Annual Review of Control, Robotics, and Autonomous Systems},
  volume       = {6},
  pages        = {447--476},
  year         = {2023},
  doi          = {10.1146/annurev-control-071822-095355}
}

@article{Slutzky2025_BCIworkingdefinition,
  author       = {Slutzky, Marc W. and Vansteensel, Mariska J. and Herff, Christian and Gaunt, Robert A.},
  title        = {A brain--computer interface working definition},
  journal      = {Nature Biomedical Engineering},
  volume       = {9},
  number       = {6},
  pages        = {792},
  year         = {2025},
  doi          = {10.1038/s41551-025-01414-8}
}

@article{Khan2020_MI_BCI_Stroke,
  author       = {Khan, Muhammad Ahmed and Das, Rig and Iversen, Helle K. and Puthusserypady, Sadasivan},
  title        = {Review on motor imagery based {BCI} systems for upper-limb post-stroke neurorehabilitation: From designing to application},
  journal      = {Computers in Biology and Medicine},
  volume       = {123},
  pages        = {103843},
  year         = {2020},
  doi          = {10.1016/j.compbiomed.2020.103843}
}

@article{Jeannerod1995_MI,
  author       = {Jeannerod, Marc},
  title        = {Mental imagery in the motor context},
  journal      = {Neuropsychologia},
  volume       = {33},
  number       = {11},
  pages        = {1419--1432},
  year         = {1995},
  doi          = {10.1016/0028-3932(95)00073-C}
}

@article{Kilner2004_MI,
  author       = {Kilner, James M. and Vargas, Catarina and Duval, Sylvain and Blakemore, Sarah-Jayne and Sirigu, Angela},
  title        = {Motor activation prior to observation of a predicted movement},
  journal      = {Nature Neuroscience},
  volume       = {7},
  number       = {12},
  pages        = {1299--1301},
  year         = {2004},
  doi          = {10.1038/nn1355}
}

@article{Lotze2006_MI,
  author       = {Lotze, Martin and Halsband, Ulrike},
  title        = {Motor imagery},
  journal      = {Journal of Physiology - Paris},
  volume       = {99},
  number       = {4–6},
  pages        = {386--395},
  year         = {2006},
  doi          = {10.1016/j.jphysparis.2006.03.012}
}

@incollection{KleihKuebler2018_UCD_BCI,
  author       = {Kleih, Sonja C. and Kübler, Andrea},
  title        = {Why user-centered design is relevant for brain--computer interfacing and how it can be implemented in study protocols},
  booktitle    = {Brain--Computer Interfaces Handbook},
  editor       = {Berger, Klaus and Chen, Lawrence and M{\"u}ller, Kai-Rainer},
  publisher    = {CRC Press, Taylor \& Francis Group},
  address      = {Boca Raton, FL},
  year         = {2018},
  pages        = {557--566},
  doi          = {10.1201/9781351231954-30}
}

@article{Tzimourta2025_HCD_DigitalHealth,
  author       = {Tzimourta, Katerina D.},
  title        = {Human-Centered Design and Development in Digital Health: Approaches, Challenges, and Emerging Trends},
  journal      = {Cureus},
  volume       = {17},
  number       = {6},
  pages        = {e85897},
  year         = {2025},
  doi          = {10.7759/cureus.85897}
}

@article{Herbert2024_LanguageCultureBCI,
  author       = {Herbert, Cornelia},
  title        = {Brain-computer interfaces and human factors: the role of language and cultural differences -- Still a missing gap?},
  journal      = {Frontiers in Human Neuroscience},
  volume       = {18},
  pages        = {1305445},
  year         = {2024},
  doi          = {10.3389/fnhum.2024.1305445}
}

@article{Gwon2023_PublicMIDatasets,
  author       = {Gwon, Daeun and Won, Kyungho and Song, Minseok and Nam, Chang S. and Jun, Sung Chan and Ahn, Minkyu},
  title        = {Review of public motor imagery and execution datasets in brain--computer interfaces},
  journal      = {Frontiers in Human Neuroscience},
  volume       = {17},
  pages        = {1134869},
  year         = {2023},
  doi          = {10.3389/fnhum.2023.1134869},
}

@article{Statthaler2017_CybathlonMirage91,
  author       = {Statthaler, Karina and Schwarz, Andreas and Steyrl, David and Kobler, Reinmar and H{\"o}ller, Maria Katharina and Brandstetter, Julia and Hehenberger, Lea and Bigga, Marvin and M{\"u}ller-Putz, Gernot},
  title        = {Cybathlon experiences of the {Graz BCI} racing team {MIRAGE91} in the brain-computer interface discipline},
  journal      = {Journal of NeuroEngineering and Rehabilitation},
  volume       = {14},
  pages        = {129},
  year         = {2017},
  doi          = {10.1186/s12984-017-0344-9}
}

@article{Neuper2001_EventRelatedDynamics,
  author       = {Neuper, Christopher and Pfurtscheller, Gert},
  title        = {Event-related dynamics of cortical rhythms: Frequency-specific features and functional correlates},
  journal      = {International Journal of Psychophysiology},
  volume       = {43},
  number       = {1},
  pages        = {41--58},
  year         = {2001},
  doi          = {10.1016/S0167-8760(01)00178-7}
}

@inproceedings{Gramfort2021SelfSupervisionEEG,
  author    = {Alexandre Gramfort and Hubert Banville and Omar Chehab and Aapo Hyv{\"a}rinen and Denis Engemann},
  title     = {Learning with self-supervision on {EEG} data},
  booktitle = {2021 9th IEEE International Winter Conference on Brain-Computer Interface ({BCI})},
  year      = {2021},
  pages     = {28--29},
  doi       = {10.1109/BCI51272.2021.9385345},
  publisher = {IEEE},
  address   = {United States}
}

@article{Delorme2004EEGLAB,
  author    = {Arnaud Delorme and Scott Makeig},
  title     = {{EEGLAB}: an open source toolbox for analysis of single-trial {EEG} dynamics including independent component analysis},
  journal   = {Journal of Neuroscience Methods},
  year      = {2004},
  volume    = {134},
  number    = {1},
  pages     = {9--21},
  doi       = {10.1016/j.jneumeth.2003.10.009}
}

@article{Ouyang2025ProtocolEEG,
  author    = {Guang Ouyang and Yingzhe Li},
  title     = {Protocol for semi-automatic {EEG} preprocessing incorporating independent component analysis and principal component analysis},
  journal   = {STAR Protocols},
  year      = {2025},
  volume    = {6},
  number    = {1},
  pages     = {103682},
  doi       = {10.1016/j.xpro.2025.103682},
  publisher = {Elsevier}
}

@article{Widmann2015DigitalFilterDesign,
  author    = {Andreas Widmann and Erich Schr{\"o}ger and Burkhard Maess},
  title     = {Digital filter design for electrophysiological data -- a practical approach},
  journal   = {Journal of Neuroscience Methods},
  year      = {2015},
  volume    = {250},
  pages     = {34--46},
  doi       = {10.1016/j.jneumeth.2014.08.002},
  publisher = {Elsevier}
}

@inproceedings{Mullen2013RealTimeModeling,
  author    = {Tim Mullen and Christian Kothe and Yu M. Chi and Alejandro Ojeda and Trevor Kerth and Scott Makeig and Gert Cauwenberghs and Tzyy-Ping Jung},
  title     = {Real-time modeling and {3D} visualization of source dynamics and connectivity using wearable {EEG}},
  booktitle = {Proceedings of the 37th Annual International Conference of the {IEEE} Engineering in Medicine and Biology Society ({EMBC})},
  year      = {2013},
  pages     = {2184--2187},
  doi       = {10.1109/EMBC.2013.6609968},
  publisher = {IEEE}
}

@article{Fatourechi2007EMGEOG,
  author    = {Mahmoud Fatourechi and Ali Bashashati and Rabab K. Ward and Gary E. Birch},
  title     = {{EMG} and {EOG} artifacts in brain computer interface systems: A survey},
  journal   = {Clinical Neurophysiology},
  year      = {2007},
  volume    = {118},
  number    = {3},
  pages     = {480--494},
  doi       = {10.1016/j.clinph.2006.10.009},
  publisher = {Elsevier}
}

@article{Ludwig2009CommonAverage,
  author    = {Karl A. Ludwig and Ryan M. Miriani and Nathan B. Langhals and Michael D. Joseph and Douglas J. Anderson and Daryl R. Kipke},
  title     = {Using a common average reference to improve cortical neuron recordings from microelectrode arrays},
  journal   = {Journal of Neurophysiology},
  year      = {2009},
  volume    = {101},
  number    = {3},
  pages     = {1679--1689},
  doi       = {10.1152/jn.91095.2008},
  publisher = {American Physiological Society}
}

@article {Tallon-Baudry722,
	author = {Tallon-Baudry, Catherine and Bertrand, Olivier and Delpuech, Claude and Pernier, Jacques},
	title = {Oscillatory {$\gamma$}-Band (30--70 Hz) Activity Induced by a Visual Search Task in Humans},
	volume = {17},
	number = {2},
	pages = {722--734},
	year = {1997},
	doi = {10.1523/JNEUROSCI.17-02-00722.1997},
	publisher = {Society for Neuroscience},
	journal = {Journal of Neuroscience}
}

@article{Koles1991QuantitativeExtraction,
  author    = {Z. J. Koles},
  title     = {The quantitative extraction and topographic mapping of the abnormal components in the clinical {EEG}},
  journal   = {Electroencephalography and Clinical Neurophysiology},
  year      = {1991},
  volume    = {79},
  number    = {6},
  pages     = {440--447},
  doi       = {10.1016/0013-4694(91)90163-X},
  pmid      = {1721571}
}

@article{MullerGerking1999OptimalSpatialFilters,
  author    = {J{\"u}rgen M{\"u}ller-Gerking and Gert Pfurtscheller and Hans Flyvbjerg},
  title     = {Designing optimal spatial filters for single-trial {EEG} classification in a movement task},
  journal   = {Clinical Neurophysiology},
  year      = {1999},
  volume    = {110},
  number    = {5},
  pages     = {787--798},
  doi       = {10.1016/S1388-2457(98)00038-8}
}

@article{GramfortEtAl2013a,
  title     = {{{MEG}} and {{EEG}} Data Analysis with {{MNE}}-{{Python}}},
  author    = {Gramfort, Alexandre and Luessi, Martin and Larson, Eric and Engemann, Denis A. and Strohmeier, Daniel and Brodbeck, Christian and Goj, Roman and Jas, Mainak and Brooks, Teon and Parkkonen, Lauri and H{\"a}m{\"a}l{\"a}inen, Matti S.},
  journal   = {Frontiers in Neuroscience},
  year      = {2013},
  volume    = {7},
  number    = {267},
  pages     = {1--13},
  doi       = {10.3389/fnins.2013.00267}
}

@article{CortesVapnik1995SupportVectorNetworks,
  author    = {Corinna Cortes and Vladimir Vapnik},
  title     = {Support-vector networks},
  journal   = {Machine Learning},
  year      = {1995},
  volume    = {20},
  pages     = {273--297},
  doi       = {10.1007/BF00994018}
}

@inproceedings{Gal2016DropoutBayesian,
  title     = {Dropout as a {Bayesian} approximation: Representing model uncertainty in deep learning},
  author    = {Yarin Gal and Zoubin Ghahramani},
  booktitle = {Proceedings of the 33rd International Conference on Machine Learning (ICML)},
  pages     = {1050--1059},
  year      = {2016},
  publisher = {JMLR}
}

@phdthesis{Jeunet2016MiBCI,
  author       = {Camille Jeunet},
  title        = {{Understanding {\&} Improving Mental-Imagery Based Brain-Computer Interface ({MiBCI}) User-Training: Towards a New Generation of Reliable, Efficient {\&} Accessible Brain-Computer Interfaces}},
  school       = {Psychology. Universit{\'e} de Bordeaux},
  year         = {2016},
  address      = {Bordeaux, France},
  note         = {NNT: 2016BORD0221. tel-01417606}
}

@article{Lotte2013FlawsBCITraining,
  author    = {Fabien Lotte and Florian Larrue and Christian M{\"u}hl},
  title     = {Flaws in current human training protocols for spontaneous Brain-Computer Interfaces: lessons learned from instructional design},
  journal   = {Frontiers in Human Neuroscience},
  volume    = {7},
  year      = {2013},
  pages     = {568},
  doi       = {10.3389/fnhum.2013.00568}
}

@article{Guger2012HowManySSVEP,
  author    = {Christoph Guger and Brendan Z. Allison and Bernhard Grosswindhager and Robert Pr{\"u}ckl and Christoph Hinterm{\"u}ller and Christoph Kapeller and Markus Bruckner and Gunther Krausz and Guenter Edlinger},
  title     = {How Many People Could Use an {SSVEP} {BCI}?},
  journal   = {Frontiers in Neuroscience},
  year      = {2012},
  volume    = {6},
  pages     = {169},
  doi       = {10.3389/fnins.2012.00169}
}

@article{Liu2022ReviewSSVEP,
  author    = {S. Liu},
  title     = {Review of Brain--Computer Interface Based on Steady-State Visual Evoked Potential ({SSVEP})},
  journal   = {Brain--Stimulation Advances},
  year      = {2022},
  volume    = {2},
  number    = {3},
  pages     = {100035},
  doi       = {10.26599/BSA.2022.9050022}
}

@article{Blankertz2008OptimizingSpatialFilters,
  author    = {Benjamin Blankertz and Ryota Tomioka and Steven Lemm and Motoaki Kawanabe and Klaus-Robert M{\"u}ller},
  title     = {Optimizing spatial filters for robust {EEG} single-trial analysis},
  journal   = {IEEE Signal Processing Magazine},
  year      = {2008},
  volume    = {25},
  number    = {1},
  pages     = {41--56},
  doi       = {10.1109/MSP.2008.4408441}
}

@article{Lawhern2018EEGNet,
  author    = {Vernon J. Lawhern and Amelia J. Solon and Nicholas R. Waytowich and Stephen M. Gordon and Chou P. Hung and Brent J. Lance},
  title     = {{EEGNet}: A compact convolutional neural network for {EEG}-based brain-computer interfaces},
  journal   = {Journal of Neural Engineering},
  year      = {2018},
  volume    = {15},
  number    = {5},
  pages     = {056013},
  doi       = {10.1088/1741-2552/aace8c}
}

@article{Liao2025EEGEncoder,
  author    = {Wangdan Liao and Weidong Wang},
  title     = {{EEGEncoder}: Advancing {BCI} with transformer-based motor imagery classification},
  journal   = {Scientific Reports},
  year      = {2025},
  doi       = {10.1038/s41598-025-06364-4}
}

@article{Tabar2016DeepLearningEEG,
  author    = {Yahya R. Tabar and Fatemeh Halici},
  title     = {A novel deep learning approach for classification of {EEG} motor imagery signals},
  journal   = {Journal of Neural Engineering},
  year      = {2016},
  volume    = {14},
  number    = {1},
  pages     = {016003},
  doi       = {10.1088/1741-2560/14/1/016003}
}

@article{Pedregosa2011ScikitLearn,
  author    = {Fabian Pedregosa and Ga{\"e}l Varoquaux and Alexandre Gramfort and Vincent Michel and Bertrand Thirion and Olivier Grisel and Mathieu Blondel and Peter Prettenhofer and Ron Weiss and Vincent Dubourg and Jake Vanderplas and Alexandre Passos and David Cournapeau and Matthieu Brucher and Matthieu Perrot and {\'E}douard Duchesnay},
  title     = {{Scikit-learn: Machine Learning in Python}},
  journal   = {Journal of Machine Learning Research},
  year      = {2011},
  volume    = {12},
  pages     = {2825--2830},
  doi       = {10.5555/1953048.2078195}
}

@article{gu2022efficientlymodelinglongsequences,
      title={{Efficiently Modeling Long Sequences with Structured State Spaces}}, 
      author={Albert Gu and Karan Goel and Christopher R\'{e}},
      year={2022},
      journal={arXiv},
      doi={https://doi.org/10.48550/arXiv.2111.00396}, 
}

@inproceedings{gu2022b,
 author = {Gu, Albert and Goel, Karan and Gupta, Ankit and R\'{e}, Christopher},
 booktitle = {Advances in Neural Information Processing Systems},
 editor = {S. Koyejo and S. Mohamed and A. Agarwal and D. Belgrave and K. Cho and A. Oh},
 pages = {35971--35983},
 publisher = {Curran Associates, Inc.},
 title = {On the Parameterization and Initialization of Diagonal State Space Models},
 url = {https://proceedings.neurips.cc/paper_files/paper/2022/file/e9a32fade47b906de908431991440f7c-Paper-Conference.pdf},
 volume = {35},
 year = {2022}
}

@inproceedings{gu2020,
 author = {Gu, Albert and Dao, Tri and Ermon, Stefano and Rudra, Atri and R\'{e}, Christopher},
 booktitle = {Advances in Neural Information Processing Systems},
 editor = {H. Larochelle and M. Ranzato and R. Hadsell and M.F. Balcan and H. Lin},
 pages = {1474--1487},
 publisher = {Curran Associates, Inc.},
 title = {HiPPO: Recurrent Memory with Optimal Polynomial Projections},
 url = {https://proceedings.neurips.cc/paper_files/paper/2020/file/102f0bb6efb3a6128a3c750dd16729be-Paper.pdf},
 volume = {33},
 year = {2020}
}

@article{gu2022trainhippostatespace,
      title={How to Train Your HiPPO: State Space Models with Generalized Orthogonal Basis Projections}, 
      author={Albert Gu and Isys Johnson and Aman Timalsina and Atri Rudra and Christopher R\'{e}},
      year={2022},
      journal={arXiv},
      doi={10.48550/arXiv.2206.12037}, 
}

@patent{asrpatent,
    inventor={Kothe, Christian Andreas Edgar and
    Jung, Tzyy-Ping},
assignee = {The Regents of the University of California, San Diego},
    title = {Artifact removal techniques with signal reconstruction},
    year = {2016},
    note = {US Patent Application No. 14/895,440, Publication No. US20160113587A1},
}

@article{10.1162/IMAG.a.136,
    author = {Kothe, Christian and Shirazi, Seyed Yahya and Stenner, Tristan and Medine, David and Boulay, Chadwick and Grivich, Matthew I. and Artoni, Fiorenzo and Mullen, Tim and Delorme, Arnaud and Makeig, Scott},
    title = {The lab streaming layer for synchronized multimodal recording},
    journal = {Imaging Neuroscience},
    volume = {3},
    pages = {IMAG.a.136},
    year = {2025},
    month = {09},
    doi = {10.1162/IMAG.a.136},
}

@article{STUCHI2024111443,
title = {A frequency-domain approach with learnable filters for image classification},
journal = {Applied Soft Computing},
volume = {155},
pages = {111443},
year = {2024},
doi = {10.1016/j.asoc.2024.111443},
author = {José Augusto Stuchi and Natalia Gil Canto and Romis Ribeiro de Faissol Attux and Levy Boccato},

}

@article{Kuebler2014,
  author  = {K{\"u}bler, Andrea and Holz, Elisa M. and Riccio, Angela and Zickler, Claudia and Kaufmann, Tobias and Kleih, Sonja C. and Staiger-S{\"a}lzer, Pit and Desideri, Lorenzo and Hoogerwerf, Evert-Jan and Mattia, Donatella},
  title   = {The User-Centered Design as Novel Perspective for Evaluating the Usability of BCI-Controlled Applications},
  journal = {PLOS ONE},
  year    = {2014},
  volume  = {9},
  number  = {12},
  pages   = {e112392},
  doi     = {10.1371/journal.pone.0112392},
}

@article{Guarnieri2018,
  author  = {Guarnieri, Roberto and Marino, Marco and Barban, Fabio and Ganzetti, Michele and Mantini, Donatella},
  title   = {Online {EEG} artifact removal for {BCI} applications by adaptive spatial filtering},
  journal = {Journal of Neural Engineering},
  year    = {2018},
  volume  = {15},
  number  = {5},
  pages   = {056005},
  doi     = {10.1088/1741-2552/aacfdf},
}

@inproceedings{Kim2013,
  author    = {Kim, Hyun Seok and Chang, Min Hye and Lee, Hong Ji and Park, Kwang Suk},
  title     = {A comparison of classification performance among the various combinations of motor imagery tasks for brain–computer interface},
  booktitle = {2013 6th International IEEE/EMBS Conference on Neural Engineering (NER 2013)},
  year      = {2013},
  pages     = {435--438},
  doi       = {10.1109/NER.2013.6695965},
}

@article{KHADEMI2023109736,
title = {A review of critical challenges in {MI-BCI}: From conventional to deep learning methods},
journal = {Journal of Neuroscience Methods},
volume = {383},
pages = {109736},
year = {2023},
doi = {10.1016/j.jneumeth.2022.109736},
author = {Zahra Khademi and Farideh Ebrahimi and Hussain Montazery Kordy}
}

@article{RAZA2019154,
title = {Covariate shift estimation based adaptive ensemble learning for handling non-stationarity in motor imagery related {EEG}-based brain-computer interface},
journal = {Neurocomputing},
volume = {343},
pages = {154--166},
year = {2019},
doi = {10.1016/j.neucom.2018.04.087},
author = {Haider Raza and Dheeraj Rathee and Shang-Ming Zhou and Hubert Cecotti and Girijesh Prasad}
}

@article{zhao2024ctnet,
  title={{CTNet: a convolutional transformer network for EEG-based motor imagery classification}},
  author={Zhao, Wei and Jiang, Xiaolu and Zhang, Baocan and Xiao, Shixiao and Weng, Sujun},
  journal={Scientific Reports},
  volume={14},
  pages={20237},
  year={2024},
  doi = {10.1038/s41598-024-71118-7}
}

@article{SADEGHI2019101607,
title = {Accurate estimation of information transfer rate based on symbol occurrence probability in brain-computer interfaces},
journal = {Biomedical Signal Processing and Control},
volume = {54},
pages = {101607},
year = {2019},
doi = {10.1016/j.bspc.2019.101607},
author = {Sahar Sadeghi and Ali Maleki}
}

@Inbook{Billinger2013,
author={Billinger, Martin
and Daly, Ian
and Kaiser, Vera
and Jin, Jing
and Allison, Brendan Z.
and M{\"u}ller-Putz, Gernot R.
and Brunner, Clemens},
editor={Allison, Brendan Z.
and Dunne, Stephen
and Leeb, Robert
and Del R. Mill{\'a}n, Jos{\'e}
and Nijholt, Anton},
chapter={Is It Significant? {Guidelines} for Reporting {BCI} Performance},
title={Towards Practical Brain-Computer Interfaces: Bridging the Gap from Research to Real-World Applications},
year={2013},
publisher={Springer Berlin Heidelberg},
address={Berlin, Heidelberg},
pages={333--354},
doi={10.1007/978-3-642-29746-5_17},
}

@Article{signals4010004,
AUTHOR = {Vavoulis, Athanasios and Figueiredo, Patricia and Vourvopoulos, Athanasios},
TITLE = {A Review of Online Classification Performance in Motor Imagery-Based Brain–Computer Interfaces for Stroke Neurorehabilitation},
JOURNAL = {Signals},
VOLUME = {4},
YEAR = {2023},
NUMBER = {1},
PAGES = {73--86},
DOI = {10.3390/signals4010004}
}

@article{JAFARIFARMAND2020101749,
title = {Real-time multiclass motor imagery brain-computer interface by modified common spatial patterns and adaptive neuro-fuzzy classifier},
journal = {Biomedical Signal Processing and Control},
volume = {57},
pages = {101749},
year = {2020},
doi = {https://doi.org/10.1016/j.bspc.2019.101749},
author = {Aysa Jafarifarmand and Mohammad Ali Badamchizadeh},
}

@ARTICLE{10.3389/fninf.2022.961089,
AUTHOR={Triana-Guzman, Nayid  and Orjuela-Cañon, Alvaro D.  and Jutinico, Andres L.  and Mendoza-Montoya, Omar  and Antelis, Javier M. },  
TITLE={Decoding {EEG} rhythms offline and online during motor imagery for standing and sitting based on a brain-computer interface},
JOURNAL={Frontiers in Neuroinformatics},
VOLUME={16},
YEAR={2022},
DOI={10.3389/fninf.2022.961089},
}

@article{10.1371/journal.pone.0268880,
    doi = {10.1371/journal.pone.0268880},
    author = {Tibrewal, Navneet AND Leeuwis, Nikki AND Alimardani, Maryam},
    journal = {PLOS ONE},
    publisher = {Public Library of Science},
    title = {Classification of motor imagery EEG using deep learning increases performance in inefficient {BCI} users},
    year = {2022},
    month = {07},
    volume = {17},
    pages = {1--18},
    number = {7},
}

@article{riemanntransferYing,
    author = {Ying, Jiahui AND Wei, Qingguo AND Zhou, Xichen},
    title = {{Riemannian geometry-based transfer learning for reducing training time in c-VEP BCIs}},
    journal = {Scientific Reports},
    year = {2022},
    volume = {12},
    pages = {9818},
    doi = {10.1038/s41598-022-14026-y}
}

@dataset{neurotum_e_v_2025_18087806,
  author       = {neuroTUM e.V.},
  title        = {neuroTUM-BCI: Cybathlon Dataset},
  month        = dec,
  year         = 2025,
  publisher    = {neuroTUM e.V.},
  version      = {1.0.0},
  doi          = {10.5281/zenodo.18087806},
  url          = {https://doi.org/10.5281/zenodo.18087806},
}

@incollection{fisherscore,
  author    = {Aggarwal, Charu C.},
  title     = {An Introduction to Data Classification},
  booktitle = {Data Classification: Algorithms and Applications},
  editor    = {Aggarwal, Charu C.},
  publisher = {CRC Press},
  address   = {Boca Raton, FL},
  year      = {2015},
  chapter   = {1},
  pages     = {4--15},
  note      = {Section 1.2 "Common Techniques in Data Classification"},
  isbn      = {978-1-4665-8674-1},
}

@article{msconv,
    author = {Zhao, Wei AND Zhang, Baocan AND Zhou, Haifeng AND Wei, Dezhi AND Huang, Chenxi AND Lan, Quan},
    title = {Multi-scale convolutional transformer network for motor imagery brain-computer interface},
    journal = {Scientific Reports},
    year = {2025},
    volume = {15},
    pages ={12935},
    doi = {10.1038/s41598-025-96611-5}
}

\end{document}